\documentclass[english]{article}
\usepackage[T1]{fontenc}
\usepackage[latin9]{inputenc}
\usepackage[letterpaper]{geometry}
\usepackage{graphicx}
\usepackage{setspace}
\usepackage{babel}

\begin{document}
\begin{singlespace}

\title{\noindent Local Charge of the $\nu=5/2$ Fractional Quantum Hall State}
\end{singlespace}

\begin{singlespace}

\author{\noindent Vivek Venkatachalam{*}, Amir Yacoby{*}, Loren Pfeiffer$^{\dagger}$,
Ken West$^{\dagger}$}
\end{singlespace}

\begin{singlespace}

\date{\noindent {*} Department of Physics, Harvard University, Cambridge,
MA, USA \\
$^{\dagger}$ Department of Electrical Engineering, Princeton
University, Princeton, NJ, USA}
\end{singlespace}
\maketitle
\begin{abstract}
\begin{singlespace}
\noindent Electrons in two dimensions and strong magnetic fields effectively
lose their kinetic energy and display exotic behavior dominated by
Coulomb forces. When the ratio of electrons to magnetic flux quanta
in the system is near $5/2$, the unique correlated phase that emerges
is predicted to be gapped with fractionally charged quasiparticles
and a ground state degeneracy that grows exponentially as these quasiparticles
are introduced. Interestingly, the only way to transform between the
many ground states would be to braid the fractional excitations around
each other, a property with applications in quantum information processing.
Here we present the first observation of localized quasiparticles
at $\nu=5/2$, confined to puddles by disorder. Using a local electrometer
to compare how quasiparticles at $\nu=5/2$ and $\nu=7/3$ charge
these puddles, we are able to extract the ratio of local charges for
these states. Averaged over several disorder configurations and samples,
we find the ratio to be $4/3$, suggesting that the local charges
are $e_{7/3}^{*}=e/3$ and $e_{5/2}^{*}=e/4$, in agreement with theoretical
predictions. This confirmation of localized $e/4$ quasiparticles
is necessary for proposed interferometry experiments to test statistics
and computational ability of the state at $\nu=5/2$.\end{singlespace}

\end{abstract}
\begin{singlespace}
\noindent When a two-dimensional electron system (2DES) is subject
to a strong perpendicular magnetic field, the physics that emerges
is controlled by interelectron Coulomb interactions. If the 2DES is
tuned such that the ratio of electrons to magnetic flux quanta in
the system ($\nu$) is near certain rational values, the electrons
condense into so-called fractional quantum Hall (FQH) phases \cite{Girvin1987}.
These strongly-correlated states are gapped and incompressible in
the bulk of the sample, but metallic and compressible along the sample
boundary, allowing current to flow around the perimeter in such a
way that the transverse conductance is precisely quantized to $G_{xy}=\nu(e^{2}/h)$.
Additionally, the electronic correlations encoded in FQH states give
rise to local excitations with a fraction of an electron charge and
braiding statistics that fall outside the conventional classification
of bosonic or fermionic. The state at $\nu=\frac{5}{2}$, unlike its
conventional odd-denominator relatives, is predicted to have the additional
property that particle interchange can evolve the system adiabatically
between orthogonal ground states \cite{Moore1991}. This property,
dubbed non-abelian braiding statistics, has been proposed as the basis
for a topological quantum computer that would be insensitive to environmental
decoherence \cite{Nayak2008,DasSarma2005}.

\noindent One necessary (but insufficient) condition for exotic braiding
statistics at $\nu=\frac{5}{2}$ is for the ground state to support
local excitations with a charge of $e_{5/2}^{*}=e/4$, where $e$
is the charge of an electron \cite{Moore1991}. Though a charge of
$e/4$ had previously been measured using shot noise techniques \cite{Dolev2008},
more recent data from the same group \cite{Dolev2010} suggest that
the value of the measured charge changes continuously as the point
contact conductance and temperature are varied, reaching an inferred
charge of unity in the weak and strong tunneling limits. Unexpected
charges have also been reported for the more conventional fractions
at 1/3, 2/3, and 7/3 \cite{Bid2009,Dolev2010}. Moreover, DC conductance
measurements in the weak tunneling regime \cite{Radu2008} suggest
a quasiparticle charge of $e_{5/2}^{*}=0.17e$, in stark contrast
to the shot noise results.

\noindent Clearly, a better understanding of the tunneling processes
that take place between quantum Hall edges in the quantum point contact
is needed in order to interpret the shot noise results. Alternatively,
one can employ a thermodynamic approach \cite{Martin2004} that probes
the quasiparticle charge in the bulk of the sample in order to infer
quasiparticle charge. Here we use a single electron transistor as
a sensitive electrometer to measure the equilibrium charge distribution
in the bulk and its dependence on the average density and magnetic
field. Our results provide clear evidence for localized charge $e/4$
quasiparticles at $\nu=5/2$.

\noindent Our measurement employs a fixed single electron transistor
(SET) as a gated device capable of sensitively measuring the local
incompressibility ($\kappa^{-1}=\frac{\partial\mu}{\partial n}$)
of a high-mobility 2DES \cite{Ilani2001}. The 2DES has a 200 nm deep,
30 nm wide MBE-grown GaAs/AlGaAs quantum well, with symmetric Si $\delta$-doping
layers 100 nm on either side. A metallic backgate grown 2 $\mu\mbox{m}$
below the 2DES allows us to tune the global density, $n$, in the
well over a typical range of $2.3-2.5\times10^{15}$m$^{-2}$, with
some variation between samples. The SET is fabricated on top of the
sample using standard electron beam lithography and shadow-evaporation
techniques (Figure 1), creating an island with dimensions 500 nm $\times$
80 nm. All measurements were carried out in a dilution refrigerator
with an electron temperature of 20 mK, verified using standard Coulomb
blockade techniques.

\noindent As we adjust the density and magnetic field we expect to
see regions of incompressibility when a gap is present, which will
only happen precisely when the system is in a QH state. The slope
of these incompressible regions in the $nB$-plane corresponds to
the filling factor of the state \cite{Ilani2004}. Figure 2 shows
incompressibility versus density and magnetic field between $\nu=2$
and $\nu=3$, with the two highlighted regions corresponding to FQH
states at $\nu=5/2$ and $\nu=7/3$.

\noindent Additionally, due to the rough disorder potential created
by remote donors, we can expect different points in space to develop
gaps at different values of the global density. Because of this, we
expect a well-developed QH state to have a percolating incompressible
region punctured by small compressible puddles which behave as either
dots or anti-dots \cite{Ilani2004}. As the global density is varied,
a given compressible puddle will occasionally be populated by quasiparticles
or quasiholes of the surrounding incompressible state. This creates
a jump in the local chemical potential, $\mu(n)$, and a spike in
the local incompressibility $\frac{\partial\mu}{\partial n}$. The
magnitude and spacing of these spikes is determined by the charging
spectrum of the puddle, which in turn is dictated by the quasiparticle
charge in the surrounding incompressible region. Namely, if the quasiparticle
charge were reduced by a factor of three for a fixed disorder potential,
we should see three times as many compressible spikes as a function
of global electron density (Figure 1 b,c).

\noindent This difference in spike frequencies has previously been
used to measure the local charge at $\nu=1/3$ and $\nu=2/3$ \cite{Martin2004}.
Unlike shot noise measurements \cite{Bid2009}, these local compressibility
measurements find a quasiparticle charge of $e/3$ at both filling
factors. Additionally, because of the spatial resolution afforded
by the scanning technique in that measurement, it was possible to
establish that the disorder potential landscape does not change as
the electron system is tuned between Hall states with comparable gaps.
Transport measurements confirm that the gap inferred from activation
of $R_{xx}$ minima is comparable for the states at 5/2 and 7/3 \cite{Dean2008a,Choi2008},
so we can expect similar potential landscapes for the two states.

\noindent Our procedure begins with obtaining charging spectra (incompressibility
versus density) at $\nu=5/2$ and $\nu=7/3$. Because the gap for
these states is comparable, and the disorder potential is not altered
as we change the magnetic field or density, we expect the spacing
between charging features to reflect the quasiparticle charge in each
state. In the limit of an isolated compressible puddle surrounded
by an incompressible fluid, this relationship is particularly simple
- if the ratio of local charges between the two spectra is $\beta$,
the spectra should be identical after one of the density axes is rescaled
by a factor of $\beta$, and shifted by some amount (Figure 3a). To
proceed, we choose a value of $\beta$ and stretch one of the spectra
by this factor. We then calculate the correlation $\left(\frac{{\langle C_{1}(x)\, C_{2}(x)\rangle}}{\sqrt{\langle C_{1}(x)^{2}\rangle\langle C_{2}(x)^{2}\rangle}}\right)$
between the two spectra as a function of density offset and record
the highest value. Finally, we repeat this for many scaling factors
to obtain quality-of-fit versus $\beta$, as depicted in Figure 3b.

\noindent This procedure was repeated for 20 different disorder configurations,
obtained by changing samples, measuring with different SETs, or thermal
cycling to change the disorder. A summary of the data is shown in
Figure 4a, with an average over the measured ensemble in Figure 4b.
The peak observed at $\beta=1.31$ suggests a charge ratio of 4:3
between the two states, and a qualitative inspection of spectra overlap
(as in Figure 3a) corroborates this. To determine the significance
of the peak value, we repeated our analysis with pairs of spectra
from different disorder configurations, which should be less correlated.
For each scale, we characterized the distribution of best correlations
with a mean and standard deviation. These, in turn, can be simply
converted to the expected mean and standard error for our data (if
it were uncorrelated). The 1$\sigma$ region around the uncorrelated
mean is depicted in red in Figure 3b. Our averaged correlation at
$\beta=1.31$ lies 3.8 standard errors above the uncorrelated mean,
corresponding to a one-tailed P-value of $7\times10^{-5}$. Assuming
a charge of $e_{7/3}^{*}=e/3$, this measured value of $\beta$ suggests
$e_{5/2}^{*}=(e/3)/(1.31)=0.254e$, in agreement with the Moore-Read
prediction of $e_{5/2}^{*}=e/4$ \cite{Moore1991}.

\noindent To better understand why some configurations seem to provide
weaker (and sometimes different) measurements of $\beta$, it helps
to abandon the assumption that we are charging and monitoring single
puddles, as well as the assumption that quasiparticles in different
puddles do not interact. A free energy for our system that takes these
into account is given by\[
F=\sum_{i}(\epsilon_{i}-V_{BG})Q_{i}+\frac{1}{2}\sum_{i}U_{i}Q_{i}(Q_{i}-1)+\sum_{i<j}V_{ij}Q_{i}Q_{j}-\sum_{i}\Delta\left\lfloor \frac{Q_{i}}{2}\right\rfloor .\]
Here, $U_{i}$ and $\epsilon_{i}$ are the onsite interaction (self-capacitance)
and bare disorder potential for puddle $i$ respectively. $V_{ij}$
is a pairwise interaction, or cross-capacitance, between puddles $i$
and $j$, and  $\Delta$ is the energy gained by forming a bound pair
of quasiparticles. For now, we will let $\Delta=0$. We assume that
some subset of the puddles is capacitively coupled to and measured
by the SET.

\noindent To compute charging spectra from this model, we first choose
values of $U,$ $V$, and $\epsilon$ for each puddle from Gaussian
distributions. We then discretize $Q_{i}$ into units of $e/3$ or
$e/4$ and determine how many units of charge to put in each puddle
to minimize the above free energy. This is done for each value of
$V_{BG}$ and converted into a charging spectrum. Finally, we can
take the resulting spectra and repeat the processing performed on
data to obtain summary statistics for comparison. The result, with
$\epsilon=0\pm.3U$ and $V_{ij}=0.3U\pm0.2U$, is shown in Figure
4c. Results for other parameter choices in a large range are qualitatively
similar, with smaller values of $\sigma_{\epsilon}$ and $V_{ij}$
corresponding to sharper peaks and less spread. As expected, these
simulations tell us that both $\epsilon$ and $V_{ij}$ can distort
spectra in such a way that the maximum cross-covariance will shift
slightly or even dramatically away from 4/3. Still, we should always
expect some weight at 4/3, and this can be extracted by averaging
over disorder configurations (Figure 4d).

\noindent Recently, there has been some suggestion that $e/2$ quasiparticles
at the $\nu=5/2$ edge may be present and relevant to interference
measurements \cite{Bishara2009}. In the context of our model, we
can consider the effect weak binding of quasiparticles would have
on measured spectra. This binding is parameterized by $\Delta$ above,
and we only consider the case where pairing affects the $e/4$ quasiparticles.
As the strength of pairing is increased relative to the onsite interaction
(Figure 4d), we expect weight to shift from the peak at $4/3$ to
a peak at $2/3$ (corresponding to $e/2$ quasiparticles), with considerable
weight at $2/3$ even when $\Delta=0.1\, U$. Our data show no appreciable
evidence for a peak at $2/3$, suggesting that the only quasiparticles
participating in localization are have charge $e/4$.

\noindent These measurements constitute the first direct measurement
of incompressibility and localized states at $\nu=5/2$, and provide
an equilibrium probe of the local charge that is insensitive to complications
that arise from measurements of transport through nanostructures.
The measured value, $e_{5/2}^{*}=e/4$, indicates that the FQH state
at $\nu=5/2$ demonstrates pairing, in agreement with proposed non-Abelian
variational wavefunctions and different from other observed FQH states.
Finally, the localization of $e/4$ quasiparticles is essential to
the development of interferometers capable of detecting and exploiting
these exotic braiding properties \cite{Stern2006,Bonderson2006},
and our measurements suggest that e/4 localization does indeed occur
in a well-behaved way.

\noindent \textbf{Acknowledgements}: We would like to acknowledge
Basile Verdene, Jonah Waissman, and Johannes Nübler for technical
assistance, and we are grateful to Bertrand Halperin for theoretical
discussions. This research has been funded by Microsoft Corporation
Project Q.

\noindent \pagebreak{}\bibliographystyle{\string"Z:/Data/Local Charging at FiveHalves/Report/naturemag\string"}
\bibliography{bibliography}

\noindent %
\begin{figure}
\centering{}\includegraphics{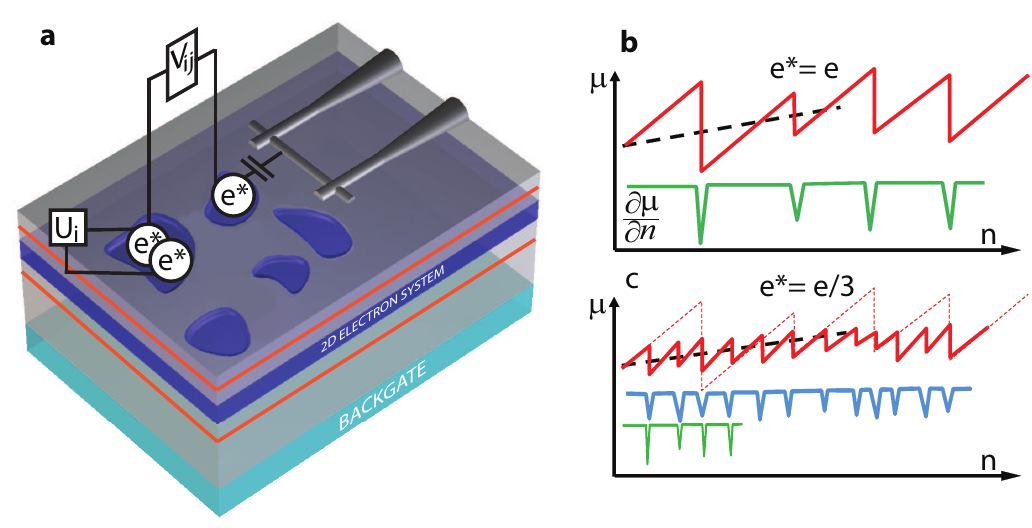}\caption{Filling puddles with fractional charge. \textbf{a}, The sample well
width is 30 nm, with symmetric Si $\delta$-doping layers 100 nm on
either side indicated by orange bands. Donors in these layers create
a disorder potential in the 2DES, which produce puddles of localized
states when the bulk is tuned to an incompressible, percolating Hall
state. These puddles have some charging energy associated with adding
electrons ($U_{i})$, and possibly some interaction with surrounding
puddles ($V_{ij}$). Incompressibility ($\kappa^{-1}=\frac{\partial\mu}{\partial n}$)
is measured using an SET fabricated on the surface. \textbf{b, }While
the global chemical potential should increase smoothly with density
(black dashed line), the local chemical potential will increase in
jumps (red line), with charge being added when the global chemical
potential aligns with a localized state. \textbf{c, }Repeating the
charging of an identical puddle with charge $e/3$ objects instead
of charge $e$ objects results in three times as many charging events
in the same range of global density. Scaling the density axis of the
charge $e$ spectrum by $1/3$ and shifting by some amount (green
curve) should result in good overlap of the incompressibility spectra. }

\end{figure}

\noindent %
\begin{figure}
\begin{centering}
\includegraphics[scale=0.8]{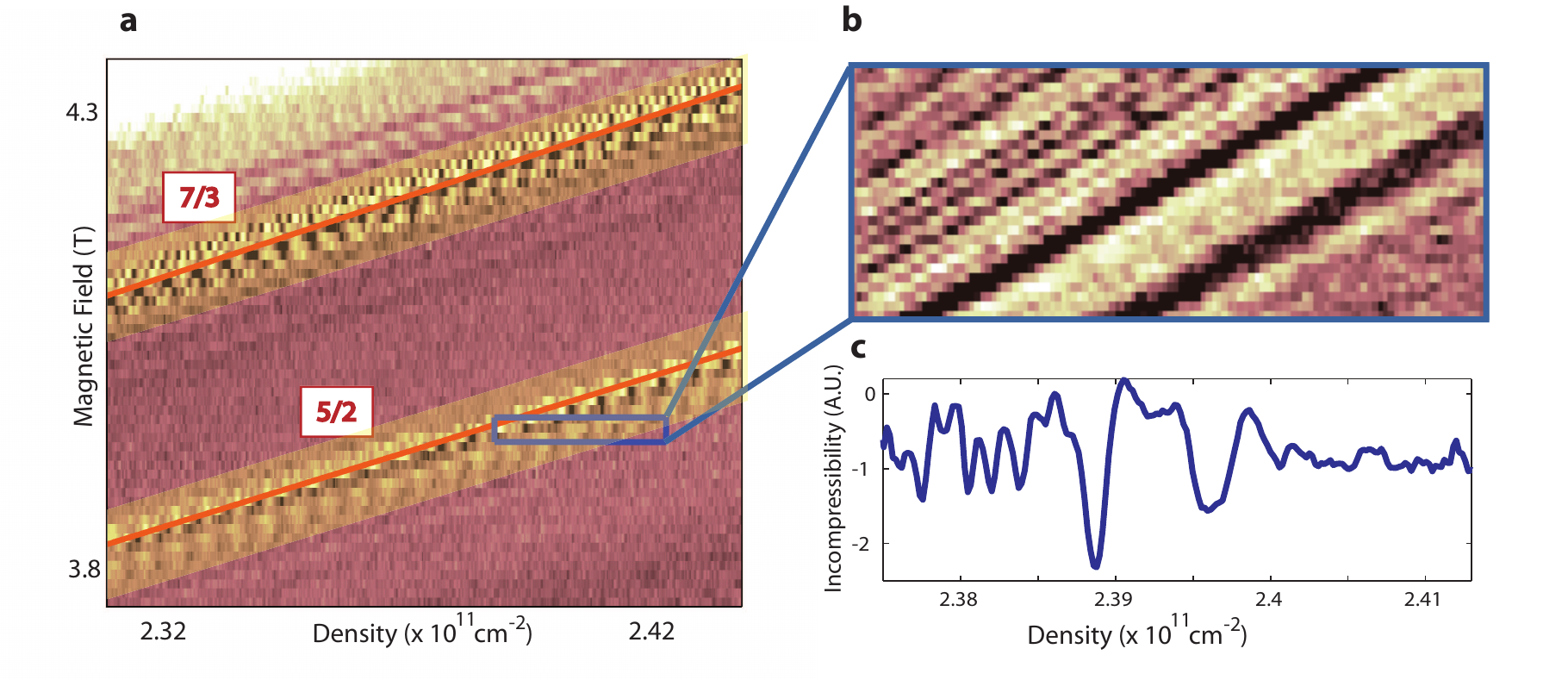}
\par\end{centering}

\caption{Incompressibility and localized states at 5/2. \textbf{a}, By varying
the magnetic field and the backgate voltage (density), we can identify
incompressible phases of the 2DES. Our samples show clear incompressible
FQH states at 5/2 and 7/3, with the expected slopes in the $nB$-plane.
\textbf{b}, Zooming in shows repeatable charging events associated
with quasiparticles localizing in puddles under the SET, stable on
a timescale of days. \textbf{c}, A linecut showing the charging spectrum
of any puddles coupled to the SET. Downwards spikes correspond to
quasiparticles entering puddles beneath the SET.}

\end{figure}

\noindent %
\begin{figure}
\begin{centering}
\includegraphics{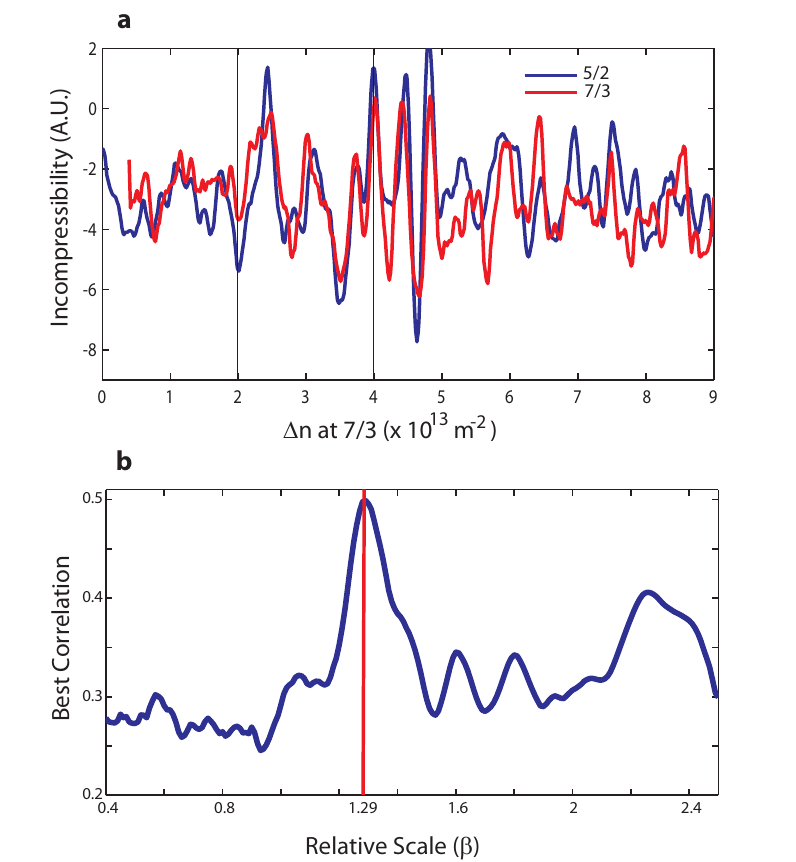}
\par\end{centering}

\caption{Comparison of spectra at 5/2 and 7/3. \textbf{a}, To determine the
charge, we first choose a relative scale between the two density axes
($\beta$), and determine the offset between the two spectra that
maximizes the cross-covariance. Here the density for the spectrum
at 5/2 is scaled up by a factor of 1.29 and shifted to match up with
the spectrum at 7/3. The guide lines show the density change required
to add 1 electron to an area of 100 nm x 500 nm, approximately the
size of our SET. We would therefore expect, very roughly, 3 e/3 charging
events in a window this size. \textbf{b}, Repeating this for many
values of $\beta$ suggests that a relative scale of 1.29 best describes
this data set.}

\end{figure}

\noindent %
\begin{figure}
\begin{centering}
\includegraphics{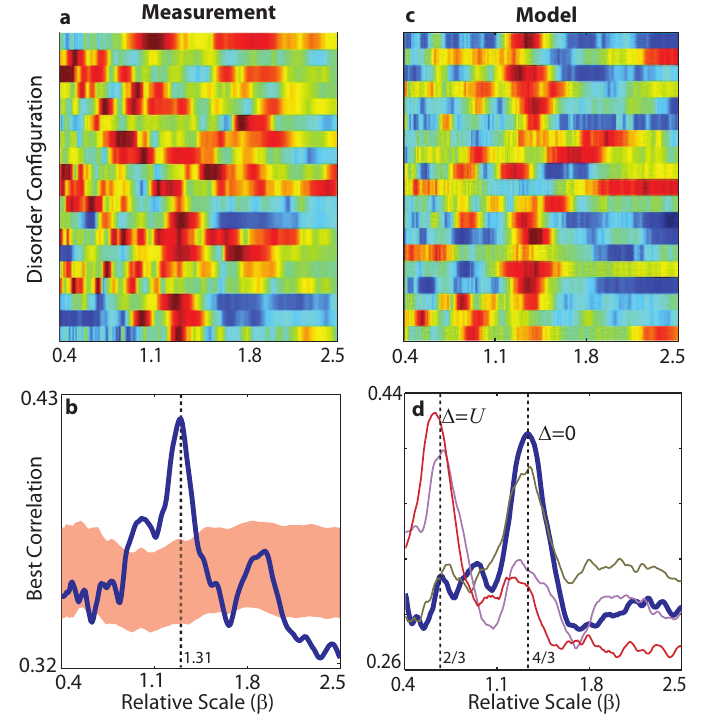}
\par\end{centering}

\caption{Summary of Data and Model. \textbf{a}, Repeating the measurement over
many disorder configurations and samples shows that the peak at 4/3
is usually present. \textbf{b}, Averaging over all measurements yields
a clear peak at $\beta=1.31$, $3.8\sigma$ above the uncorrelated
background for that scale ($P=7\times10^{-5}$), suggesting a local
charge ratio of 4/3. \textbf{c, d}, Running our model with parameters
$\epsilon=0\pm.3,\; V=0.3\pm0.2,$ and $\Delta_{5/2}=0.01,\:0.1,$
and $1.0$ (all in units of $U$, the on-site charging energy). We
simulated charging of four puddles, of which two were capacitively
coupled to the SET.}

\end{figure}
\end{singlespace}

\end{document}